\documentclass[lettersize,journal]{IEEEtran}

\usepackage{mathtools,amssymb,amsfonts}
\usepackage{algorithmic}
\usepackage{algorithm}
\usepackage{array}
\usepackage[caption=false,font=normalsize,labelfont=sf,textfont=sf]{subfig}

\usepackage{textcomp}
\usepackage{url}
\usepackage{verbatim}
\usepackage{graphicx}
\usepackage{cite}
\usepackage{tikz}
\usepackage{xcolor}
\usepackage{amsthm}
\usepackage{cuted}
\usepackage{stfloats}
\usepackage{capt-of}  
\usepackage{cuted} 

\usepackage{soul}

\usepackage[colorlinks=true,
            linkcolor=blue,
            urlcolor=blue,
            citecolor=blue]{hyperref} 
\begin{document}

\title{Optimal Bit Detection in Thermal Noise Communication Systems Under Rician Fading}

\author{Mohamed El Jbari, Fernando D. A. García, Hugerles S. Silva, Felipe A. P. de Figueiredo, \\ and Rausley A. A. de Souza
\thanks{This work has been partially funded by the xGMobile Project (XGM-AFCCT-2024-9-1-1 and XGM-AFCCT-2025-8-1-1) with resources from EMBRAPII/MCTI (Grant 052/2023 PPI IoT/Manufatura 4.0), by CNPq (302085/2025-4, 306199/2025-4), and FAPEMIG (APQ-03162-24, PPE-00124-23, RED-00194-23).}
\thanks{ Mohamed El Jbari is with LabTIC Laboratory, National School of Applied Sciences of Tangier (ENSAT), Abdelmalek Essaadi University, Morocco, e-mail: Mohamed.eljbari2@etu.uae.ac.ma.}
\thanks{Fernando D. A. García, Felipe A. P. de Figueiredo, and Rausley A. A. de Souza are with the National Institute of Telecommunications (Inatel), Santa Rita do Sapucaí, Minas Gerais, Brazil, e-mail: [fernando.almeida, felipe.figueiredo, rausley]@inatel.br.}
\thanks{Hugerles S. Silva is with the Electrical Engineering Department, University of Brasília, Federal District, Brazil, e-mail: hugerles.silva@av.it.pt.}
\vspace{-0.95cm}
}



\maketitle

\begin{abstract}
Thermal noise communication (TNC) enables ultra-low-power wireless links for Internet of Things (IoT) devices by modulating the variance of thermal noise, rather than using active carriers. 
Existing analyses often rely on Gaussian approximations and overlook fading effects, which limits their accuracy. 
This paper presents an accurate analytical framework for optimal bit detection in TNC systems under Rician fading.
Using chi-squared statistics, we derive the optimal maximum-likelihood detection threshold and an expression for the bit error probability (BEP) via Gauss–Laguerre quadrature. 
The proposed model eliminates approximation errors and accurately characterizes performance for finite sample sizes. 
Monte Carlo simulations confirm the analytical results and demonstrate significant improvements in BEP compared with suboptimal Gaussian-based detection. 
Furthermore, the influence of key parameters, sample size, resistance ratio, and Rician $K$-factor, is quantified. 
The proposed framework provides a solid foundation for designing energy-efficient TNC receivers in future B5G/6G and large-scale IoT systems.
\end{abstract}


\section{Introduction}\label{Sec.I}

\IEEEPARstart{T}{he} proliferation of energy-constrained devices in beyond five generation (B5G) and Internet of Things (IoT) networks necessitates modulation schemes that minimize power consumption without compromising performance \cite{IoT}. 
Traditional active communication systems, reliant on amplitude or phase modulation, face inherent limitations in ultra-low-power applications. 
In this context, the concept of thermal noise modulation~(TNM) has recently emerged \cite{QCN,kapetanovic2022communication}. 
TNM addresses the challenge by exploiting the variance of thermal noise to encode information, thereby eliminating the need for dedicated radio frequency carriers. 
This approach enables microwatt-level transmission, making TNM a cornerstone for sustainable and large-scale IoT deployments.

In the literature, prior studies on TNM have focused on hardware implementations \cite{kapetanovic2022communication} and performance analyses under Gaussian assumptions \cite{basar2023communication, basar2024noise}. 
For instance, bit error probability (BEP) approximations using the \(Q\)-function are derived in~\cite{basar2023communication, basar2024noise}, which is valid only for large sample sizes, while \cite{basar2024noise, 11107228} extended this to a non-coherent TNM system under Rayleigh fading channels. 
However, these approximations introduce significant inaccuracies in threshold detection, BEP calculation, and their optimization, particularly for practical sample counts. 
This occurs since the Gaussian assumption in those works neglects the chi-squared nature of the variance estimator, leading to sub-optimal results and unreliable performance predictions. 
In \cite{TNMOptDet}, a maximum likelihood detector is used to calculate the optimal threshold value for detection under constant static channels, and a closed-form BEP expression is derived considering a binary modulated thermal noise communication (TNC) system. 
Despite recent studies on TNM, existing works lack rigorous analysis of their behavior and parameter trade-offs under realistic fading conditions, such as Rician channels, which encompass both line-of-sight (LoS) and scattered components.

In this context, for the first time and aligned with the stringent power constraints of IoT, 5G, and 6G networks, we characterize binary TNC systems under Rician fading channels.
Our contributions are threefold: i) To avoid Gaussian approximation effects, we use chi-squared statistics to determine the exact choice of maximum likelihood detection (MLD) rule for the optimum bit detection threshold and the achieved BEP performance under Rician fading channels; ii) A novel BEP expression under Rician fading channels is deduced, validated against Monte-Carlo simulations, and the impact of the \(K\)-factor (ratio of direct to diffuse signal power) on the system performance is evaluated; and iii) We validate how the resistance ratio \(\alpha\), signal-to-noise~(SNR) variance ratio \(\delta\) (i.e., a measurement akin to the SNR in conventional communication systems), and sample count \(N\) jointly influence the BEP under constant static and Rician fading channels.
Our study focuses on precisely characterizing TNM as a reliable and scalable solution for next-generation low-power networks, while providing a mathematical foundation for future extensions to multi-user and dynamic fading scenarios. 
Simulation results demonstrate the advantages of the proposed MLD-based optimal bit detection for TNC systems operating in fading channels.

\textit{Notation}: $f_{(\cdot)} (\cdot)$ denotes probability density function (PDF); $F_{(\cdot)} (\cdot)$, cumulative distribution function (CDF); $\Pr(\cdot)$, probability; $I_0(\cdot)$, the zeroth-order modified Bessel function of the first kind; $\gamma_{\text{inc}}(\cdot, \cdot)$, the lower incomplete gamma function; $\mathcal{CN}(a, b)$, a circularly symmetric complex Gaussian distribution with mean $a$ and variance $b$; $\text{Exp}(a)$, an exponential distribution with rate parameter $a$; $\chi_a^2$, a central chi-squared distribution with $a$ degrees of freedom; $\mathcal{G} (a,b)$, a Gamma distribution with shape $a$ and scale $b$; $Q_{1}(a,b)$ is generalized Marcum Q-function of order 1, and ``$\sim$'' indicates ``distributed as''.

\section{System Model}\label{Sec.II}

\begin{figure}[t!]
   \centering
   \includegraphics[scale=0.365]{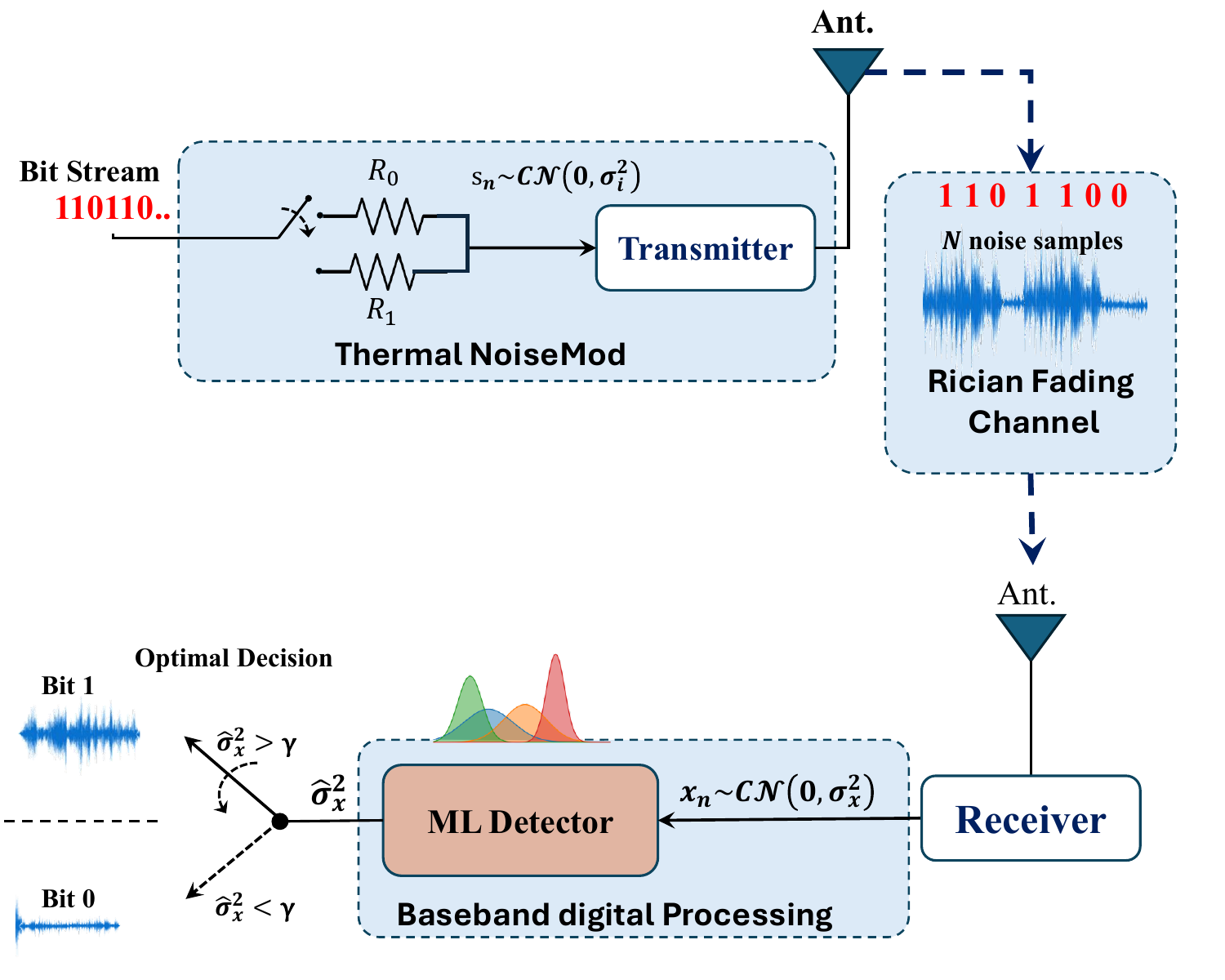}
   \vspace{-0.7cm}
   \caption{Thermal noise communication system. }
    \label{Fig.1}
   \vspace{-0.5cm}
\end{figure}

In this section, we briefly present the TNM principle for the TNC system as introduced in \cite{basar2023communication}. 
Fig.~\ref{Fig.1} presents a general transceiver structure of a TNM-based wireless communication system under Rician fading. 

At the transmitter, two levels of noise samples are generated for each bit within the symbol duration. 
The noise samples have variance given by $\sigma_i^2 = 4 R_i k T B$, $i \in \{0, 1\}$, where $R_i$ denotes the selected resistor corresponding to the input bit~\cite{basar2023communication}. 
Here, $k$, $T$, and $B$ represent the Boltzmann constant, the absolute temperature (in Kelvins), and the signal bandwidth, respectively.
In turn, the $n$-th received complex baseband sample is given by
    \begin{equation}
        x_n = h \, s_n + w_n,
    \label{eq1}\end{equation}
where $n \in \left\{1,2,\hdots,N \right\}$, $N$ denotes the total number of collected samples during the symbol duration, and $h$ is the complex channel coefficient, \( w_n \sim \mathcal{CN}(0, \sigma_w^2) \) denotes the additive white Gaussian noise (AWGN), and $ s_n \sim \mathcal{CN}(0, \sigma_i^2)$, $i  \in \left\{ 0,1\right\}$, is the transmitted noise sample with variance $\sigma_0^2$, for bit 0, or $\sigma_1^2$, for bit 1, 
where \( \sigma_w^2< \sigma_0^2 < \sigma_1^2 \). 
In this work, two scenarios for $h$ are considered. 
In the first one, it is assumed to be known due to perfect channel estimation within a coherence interval that spans one or more symbols. 
In the second scenario, the channel envelope, $|h|$, is assumed to follow a Rician distribution across different coherence intervals.

Assuming that $s_n$ and $w_n$ are circularly symmetric complex Gaussian random variables, it follows that, conditioned on $h$, the set $\left\{x_n\right\}_{n=1}^{N}$ consists of independent and identically distributed (i.i.d.) samples, each following $x_n | h \sim \mathcal{CN}(0, \sigma_{x_i}^2)$, where 
 \begin{equation}
\sigma_{x_i}^2 = \begin{cases} 
\sigma_{x_0}^2 = |h|^2 \, \sigma_0^2 + \sigma_w^2, & \text{for bit 0}, \\ 
\sigma_{x_1}^2 = |h|^2 \, \sigma_1^2 + \sigma_w^2, & \text{for bit 1}.
\end{cases}\label{eq3}
\end{equation}

The ratio between the noise variances $\sigma_{x_1}^2$ and $\sigma_{x_0}^2$ is controlled by a predefined parameter $\alpha$, such that $\alpha = \sigma_{x_1}^2 / \sigma_{x_0}^2 = R_1 / R_0$. 
Furthermore, $\delta$ is defined as the ratio between the variance of the useful (information-carrying) noise $\sigma_{x_0}^2$ and that of the receiver (interfering) noise $\sigma_w^2$, i.e., $\delta = \sigma_{x_0}^2 / \sigma_w^2$, which serves as a performance metric, such as the SNR in digital communication systems~\cite{basar2024noise, TNMOptDet}.

\section{TNC-Based Bit Detection}\label{Sec.III}

Let $\mathbf{x} = [x_1, x_2, \ldots, x_{N}] \in \mathbb{C}^N$ denote the vector containing the $N$ collected complex noise samples of a symbol, as detailed in \eqref{eq1}.
Correspondingly, 
$\mathbf{y} = [y_1, y_2, \ldots, y_{2N}] \in \mathbb{R}^{2N}$ represents the real-valued counterpart of $\mathbf{x}$, obtained by decomposing each complex sample into its real and imaginary parts.
Specifically, the first $N$ elements of $\mathbf{y}$ correspond to the real parts of $\mathbf{x}$, while the remaining $N$ elements correspond to the imaginary parts.
Conditioned on $h$, the sequence $\{y_n\}_{n=1}^{2N}$ consists of i.i.d.\ real-valued samples, each distributed as $y_n | h \sim \mathcal{N}(0,\sigma_{x_i}^2/2)$ since the real and imaginary components of $x_n$ equally share the total variance.

The conditioned PDF of $\mathbf{y}$ given the variance of the transmitted bit, $\sigma_i^2$, and $h$ is expressed as
\begin{equation}
f_\mathbf{Y}\left(\mathbf{y}|\sigma_i^2 ,h \right)=\prod_{n=1}^{2N} \frac{1}{\sqrt{ \pi \sigma_{x_i}^2}} \exp \left(-\frac{y_n^2}{ \sigma_{x_i}^2}\right). \label{eq: joint PDF y} 
\end{equation}

The MLD rule, denoted by $\mathcal{L}$, for symbol variance detection can be expressed as
\begin{equation}
    \mathcal{L} \triangleq 
    \frac{p(\mathbf{y} \,|\, \sigma_0^2, h)}{p(\mathbf{y} \,|\, \sigma_1^2, h)} 
    \underset{\sigma_0^2}{\overset{\sigma_1^2}{\lessgtr}} 
    \frac{\Pr(\sigma_1^2)}{\Pr(\sigma_0^2)},
    \label{eq5}
\end{equation}
where $\Pr(\sigma_0^2)$ and  $\Pr(\sigma_1^2)$ are the a priori probabilities of sending bit 1 and bit 0, respectively.
By substituting \eqref{eq: joint PDF y} into \eqref{eq5}, assuming that information bits are equally probable, i.e., $\Pr(\sigma_0^2) = \Pr(\sigma_1^2)$, taking the natural logarithm of both sides, and performing some algebraic simplifications, the MLD rule can be rewritten as
\begin{equation}
    \hat{\sigma}_x^2 
    \underset{\sigma_1^2}{\overset{\sigma_0^2}{\lessgtr}} 
    \underbrace{\frac{\sigma_{x_0}^2 \sigma_{x_1}^2}{\sigma_{x_1}^2 - \sigma_{x_0}^2}}_{\text{amplitude scaling}}
    \times
    \underbrace{\ln\!\left(\frac{\sigma_{x_1}^2}{\sigma_{x_0}^2}\right)}_{\text{logarithmic term}}
    \triangleq \gamma,
    \label{eq6}
\end{equation}
where $\hat{\sigma}_x^2$ denotes the estimated sample variance of the received waveform, computed from the $N$ collected samples within one symbol period, i.e., \(\hat{\sigma}_x^2 = \frac{1}{N} \sum_{n=1}^{N} |x_n|^2 = \frac{1}{N} \sum_{n=1}^{2N} y_n^2\)~\cite{CHannelEstimationTNM}, and $\gamma$ is the detection threshold.
Accordingly, the detector decides in favor of bit~0 if $\hat{\sigma}_x^2 < \gamma$, and in favor of bit~1 if $\hat{\sigma}_x^2 > \gamma$. 
Based on \eqref{eq3} and using the fact that \({ \sigma_{x_0}^2}=\delta\sigma_w^2\) and \({ \sigma_{x_1}^2}=\alpha\delta\sigma_w^2\), the optimal detection threshold can be expressed as a function of $|h|$ as 
\begin{equation}\label{eq8}
    \gamma\left(|h|\right) = \frac{\sigma_w^2 ({|h|^2}  \delta + 1)({|h|^2}  \alpha \delta + 1)}{{|h|^2}  \delta (\alpha - 1)} \ln\left( \frac{{|h|^2}  \alpha \delta + 1}{{|h|^2}  \delta + 1} \right).
\end{equation}
The conditioned BEP under optimal bit detection at the TNC receiver can be found as
    \begin{IEEEeqnarray}{lcl}\label{eq:BEPMEDIA1}
       P_{\text{e}}(|h|){=} \frac{\left[ \Pr({\hat{\sigma}_x^2} > \gamma\left(|h|\right)| \sigma_0^2 {,h}) {+} \Pr({\hat{\sigma}_x^2} < \gamma\left(|h|\right)|\sigma_1^2 {,h} ) \right]}{2} ,\IEEEeqnarraynumspace
    \end{IEEEeqnarray}
where \( \Pr({\hat{\sigma}_x^2} > \gamma\left(|h|\right)|\sigma_0^2{,h}) \) denotes the probability of deciding for bit~1 when bit~0 is transmitted, and \( \Pr({\hat{\sigma}_x^2} < \gamma\left(|h|\right) | \sigma_1^2{,h}) \) represents the probability of deciding for bit~0 when bit~1 is transmitted.

\subsection{TNC under Constant Channel Fading}\label{SubSec.I}

In this case, {due to full channel state information (CSI) during a coherence interval}, $h$ is assumed to be a known {deterministic} complex coefficient.
{
Thus, each sample $x_n$ follows a circularly symmetric complex Gaussian distribution, such that $x_n\sim\mathcal{CN}(0, \sigma_{x_0}^2)$, when bit~0 is transmitted, and $x_n \sim \mathcal{CN}(0, \sigma_{x_1}^2)$, when bit~1 is transmitted.}

{According to \eqref{eq:BEPMEDIA1}, to obtain the BEP, we first need to determine the statistics of the estimated sample variance 
$\hat{\sigma}_{x|i,h}^2 = \frac{1}{N} \sum_{n=1}^{N} |x_{n|i,h}|^2$, where $x_{n|i,h}$ is the conditional random variable of $x_n$ given $h$ and that $s_i, i \in \{0, 1\} $ was transmitted, such that $x_{n|i,h} \sim \mathcal{CN}(0, \sigma_{x_i}^2)$. However, for the sake of conciseness, we use $\hat{\sigma}_{x}^2$ and $x_{n}$ henceforth. Since $|x_{n}|^2 \sim \mathrm{Exp}\!\left(\tfrac{1}{\sigma_{x_i}^2}\right)$, it follows that 
$S \triangleq \sum_{n=1}^{N} |x_{n}|^2$ is the sum of $N$ independent exponential random variables with a common rate parameter ${1}/{\sigma_{x_i}^2}$, 
implying that $S \sim \mathcal{G}(N, \sigma_{x_i}^2)$ and consequently 
$\hat{\sigma}_{x}^2 \sim \mathcal{G}\!\left(N, {\sigma_{x_i}^2}/{N}\right)$.
For convenience, let $Z \triangleq {2N\hat{\sigma}_{x}^2}/{\sigma_{x_i}^2}$. 
After a standard change of variables, it follows that $Z \sim \chi^2_{2N}$, with  CDF given by
\begin{IEEEeqnarray}{lcl}
    \label{eq: CDF Z}
    F_{Z}(z) = \frac{\gamma_{\text{inc}}\!\left(N, \tfrac{z}{2}\right)}{\Gamma(N)}.
\IEEEeqnarraynumspace
\end{IEEEeqnarray}

The conditional probabilities required in \eqref{eq:BEPMEDIA1} can be directly obtained from the CDF in \eqref{eq: CDF Z} as follows}
\begin{align}\label{eq:PROB1}
\Pr({\hat{\sigma}_{x}^2} > \gamma\left(|h|\right) | {\sigma_0^2, h}) & = \Pr\left( \frac{2N {\hat{\sigma}_{x}^2}}{\sigma_{x_0}^2} > \frac{2N \gamma\left(|h|\right)}{\sigma_{x_0}^2} \Bigg| {\sigma_0^2,h} \right)\nonumber\\ &= 1 - F_{{Z}} \left( \frac{2N \gamma\left(|h|\right)}{\sigma_{x_0}^2} \right)\\ \label{eq:PROB2}
\Pr({\hat{\sigma}_x^2} < \gamma\left(|h|\right) | {\sigma_1^2,h}) &= \Pr\left( \frac{2N {\hat{\sigma}_x^2}}{\sigma_{x_1}^2} < \frac{2N \gamma\left(|h|\right)}{\sigma_{x_1}^2}\Bigg| {\sigma_1^2,h} \right) \nonumber\\&= F_{{Z}} \left( \frac{2N \gamma\left(|h|\right)}{\sigma_{x_1}^2} \right).
\end{align}

Plugging \eqref{eq:PROB1} and \eqref{eq:PROB2} into \eqref{eq:BEPMEDIA1}, the BEP conditioned on $h$ can be expressed as
\begin{IEEEeqnarray}{lcl}\label{eq:BEP2}
P_{\text{e}}(|h|) = \frac{1}{2} \left[ 1 - F_{Z} \left( \frac{2 N \gamma(|h|)}{\sigma_{x_0}^2} \right) + F_{Z} \left( \frac{2N \gamma(|h|)}{\sigma_{x_1}^2} \right) \right].\IEEEeqnarraynumspace
\end{IEEEeqnarray}
{The BEP can also be written as a function of the TNC parameters \(\delta\) and \(\alpha\) as}
\begin{IEEEeqnarray}{lcl}\label{eq15}
\nonumber && P_{\text{e}}(|h|)= \\
&& \   \frac{1}{2} \left[ 1 - F_{{Z}} \left( \frac{2N \gamma(|h|) }{\sigma_w^2 (1 + |h|^2\delta)} \right)  +   F_{{Z}} \left( \frac{2N \gamma(|h|) }{\sigma_w^2 (1 + |h|^2\alpha\delta )}\right)\right].\IEEEeqnarraynumspace
\end{IEEEeqnarray}

\subsection{TNC under Rician Fading}\label{SubSec.II}
In this case, {the channel envelope of $h$, $R \triangleq |h|$, randomly changes along different coherence intervals, following a Rician distribution, whose PDF is given by}
\begin{equation}
   f_{R}(r) = \frac{r}{\sigma_\text{s}^2} \exp\left( -\frac{r^2 + \lambda^2}{2\sigma_\text{s}^2} \right) I_0\left( \frac{\lambda r }{\sigma_\text{s}^2} \right), \label{PDF}
\end{equation}
where $\lambda$ denotes the amplitude of the LoS component, and $\sigma_\text{s}^2$ represents the scattered power per quadrature component. 
The Rician $K$-factor is defined as $K = \lambda^2 / (2\sigma_\text{s}^2)$, such that 
$\lambda^2 = {K}/{(K+1)}$ and $2\sigma_\text{s}^2 = {1}/{(K+1)}$.

The unconditional BEP over the Rician {fading channels can be derived from the constant channel case as 
\begin{align}
    \label{eq: BEP Rice def}
    P_{\text{e}}= \int_{0}^{\infty} P_{\text{e}}(r) f_{R}(r) \text{d}r.
\end{align}

Substituting \eqref{eq15} and \eqref{PDF}  into \eqref{eq: BEP Rice def}, we obtain}  
\begin{IEEEeqnarray}{lcl}\label{eqR1}
        &&P_{\text{e}} 
    = \int_0^\infty \dfrac{r}{2\sigma_\text{s}^2} I_0\left(\dfrac{r\lambda}{\sigma_\text{s}^2}\right)\exp\left(-\dfrac{r^2 + \lambda^2}{2\sigma_\text{s}^2}\right)\nonumber \\
    && \times\left[ 1 - F_{{Z}} \left( \frac{2N \gamma(r)}{\sigma_w^2 (1 + r^2\delta )} \right)  +   F_{{Z}} \left( \frac{2N \gamma(r)}{\sigma_w^2 (1 + r^2\alpha\delta)}\right)\right] \text{d}r.
\IEEEeqnarraynumspace
\end{IEEEeqnarray}
{After performing the change of variables} \( x = {r^2}/{(2\sigma_\text{s}^2)} \), so that \(r=\sqrt{2\sigma_\text{s}^2 x}\) and \( \text{d}r = \sqrt{{\sigma_\text{s}^2}{(2x)}} \text{d}x \), the unconditional BEP becomes
\begin{equation}
    P_{\text{e}}  = \dfrac{1}{2}  e^{-\lambda^2/(2\sigma_\text{s}^2)} \left[I_1-I_2 + I_3\right],\label{eqG}
\end{equation}
where
\begin{subequations}\label{eq:Ix}
	\begin{IEEEeqnarray}{lcl}\label{eq:I1}
  I_1 =  \int_0^\infty e^{-x} I_0\left(\lambda\sqrt{\frac{2x}{\sigma_\text{s}^2}}\right)\text{d}x,\\
  I_2 {=}  \int_0^\infty e^{-x} I_0\left(\lambda\sqrt{\frac{2x}{\sigma_\text{s}^2}}\right) F_{{Z}}\left(\frac{2 {N} \gamma(\sqrt{2\sigma_\text{s}^2 x})}{\sigma_w^2\left(1 + 2\sigma_\text{s}^2 x \delta\right)}\right)\text{d}x,\label{I2}\\
\noalign{\noindent and
	\vspace{2\jot}}
  I_3 {=}\int_0^\infty e^{-x} I_0\left(\lambda\sqrt{\frac{2x}{\sigma_\text{s}^2}}\right) F_{{Z}}\left(\frac{2 {N} \gamma(\sqrt{2\sigma_\text{s}^2 x})}{\sigma_w^2\left(1 + 2\sigma_\text{s}^2 x \alpha\delta\right)}\right)\text{d}x.\label{I3}\IEEEeqnarraynumspace
\end{IEEEeqnarray}\end{subequations}
To simplify the calculation of $I_1$, $I_2$, and $I_3$, which is convoluted, we use the Gauss-Laguerre quadrature method~\cite{press2007numerical}. 
This technique enables numerical approximation of integral expressions, significantly reducing the computational cost of evaluation. 
Thus \eqref{eq:Ix} can be rewritten as
\begin{subequations}\label{eq:Ixsolved}
	\begin{IEEEeqnarray}{lcl}
I_1 \approx  \sum_{j=1}^{N_a} w_j I_0\left(\lambda\sqrt{\dfrac{2x_j}{\sigma_\text{s}^2}}\right),\label{eqI11}\\
I_2 \approx \sum_{j=1}^{N_a} w_j I_0\left(\lambda\sqrt{\dfrac{2x_j}{\sigma_\text{s}^2}}\right) F_{{Z}}\left(\frac{2 {N} \gamma(\sqrt{2\sigma_\text{s}^2 x_j})}{\sigma_w^2\left(2\sigma_\text{s}^2 x_j \delta + 1\right)}\right),\\
\noalign{\noindent and
	\vspace{2\jot}}
I_3 \approx  \sum_{j=1}^{N_a} w_j I_0\left(\lambda\sqrt{\dfrac{2x_j}{\sigma_\text{s}^2}}\right) F_{{Z}}\left(\frac{2 {N} \gamma(\sqrt{2\sigma_\text{s}^2 x_j})}{\sigma_w^2\left(2\sigma_\text{s}^2 x_j\alpha\delta + 1\right)}\right),\label{eqI33}
\IEEEeqnarraynumspace
\end{IEEEeqnarray}\end{subequations}
where $x_j$ denotes the roots of the Laguerre polynomial $L_{N_a}(x)$, $w_j$ are the corresponding weights, and $N_a$ denotes the order of the quadrature rule, i.e., the number of sampling points used to approximate the integral.

{Replacing \eqref{eq:Ixsolved} into \eqref{eqG}, and after some simplifications, we obtain the average BEP of the TNC system under {Rician fading as}}
\begin{IEEEeqnarray}{lcl}\label{eq25}
P_{\text{e}}  \approx&& \frac{e^{-\lambda^2/(2\sigma_\text{s}^2)}}{2}   \sum_{j=1}^{N_a} w_jI_0\left(\lambda\sqrt{\dfrac{2x_j}{\sigma_\text{s}^2}}\right)\nonumber\\  &\times&\Bigg[1-F_{{Z}}\left(\frac{2N\gamma(\sqrt{2\sigma_\text{s}^2 x_j})}{\sigma_w^2\left(2\sigma_\text{s}^2 x_j \delta + 1\right)}\right)+ \nonumber\\  &&F_{{Z}}\left(\frac{2N\gamma(\sqrt{2\sigma_\text{s}^2 x_j})}{\sigma_w^2\left(2\sigma_\text{s}^2 x_j \alpha\delta + 1\right)}\right)\Bigg].\IEEEeqnarraynumspace
\end{IEEEeqnarray}
\begin{figure*}[ht]
    \centering
    \subfloat[]{
        \includegraphics[trim=0cm 0cm 0cm 0cm, clip, width=0.325\textwidth]{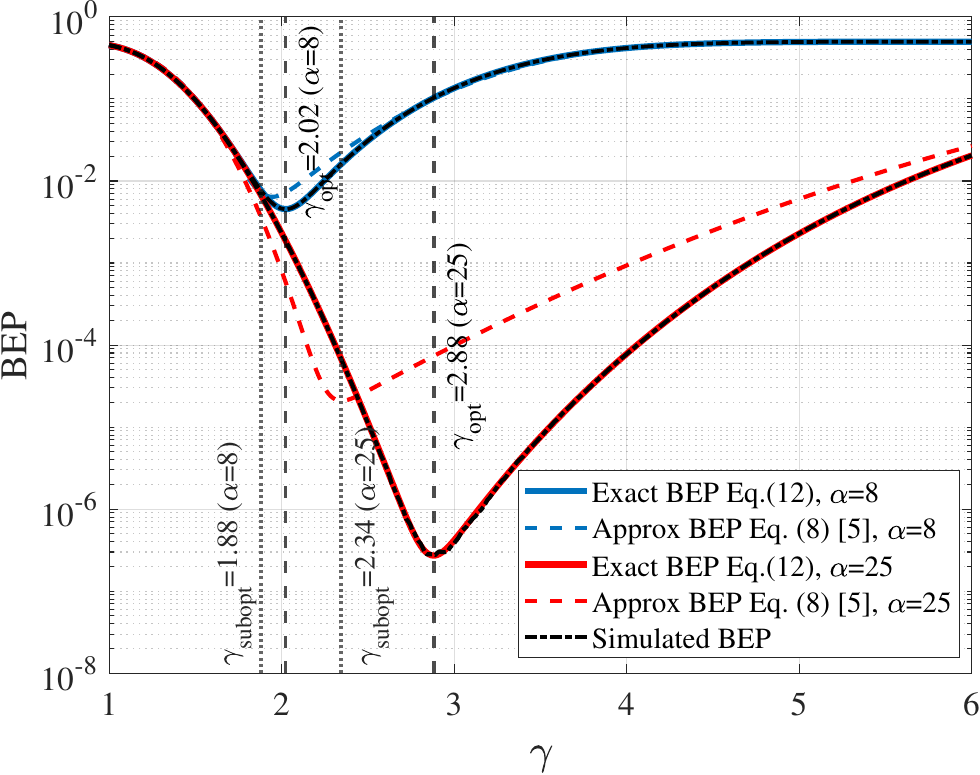}
        \label{fig:2a}
    }
    \subfloat[]{
        \includegraphics[trim=0cm 0cm 0cm 0cm, clip, width=0.325\textwidth]{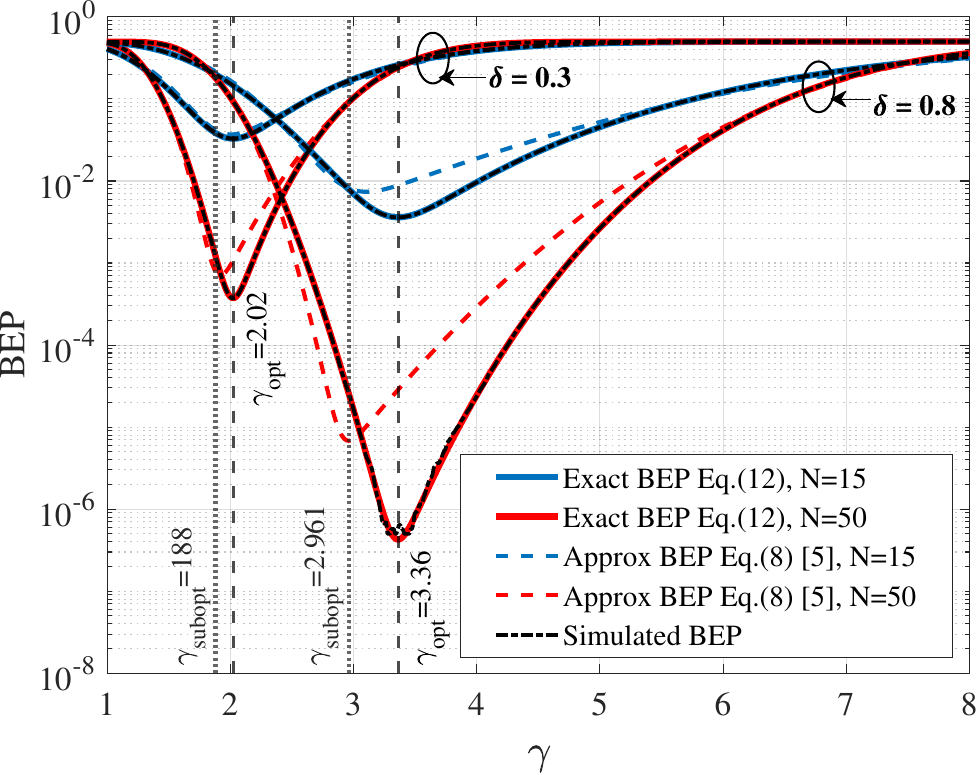}
        \label{fig:2b}
    }
    \subfloat[]{
        \includegraphics[trim=0cm 0cm 0cm 0cm, clip, width=0.325\textwidth]{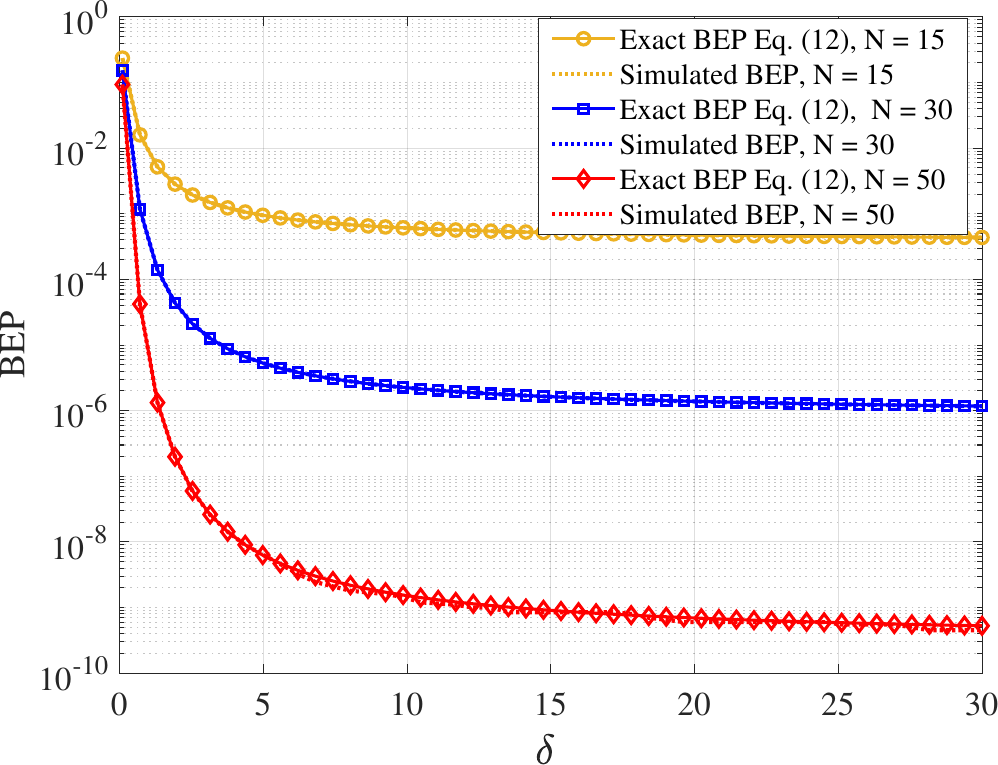}
        \label{fig:2c}
    }
    \caption{BEP performance for constant channel ($h=1$) versus:  
    (a) threshold $\gamma$ for $\delta=0.3$, $N=30$, and $\alpha=\{8,25\}$;  
    (b) threshold $\gamma$ for $\alpha=8$ and $N=\{15,50\}$;  
    (c) $\delta$ for $\alpha=8$ and $N=\{15,30,50\}$.  
    The vertical lines indicate optimal and suboptimal thresholds.}
    \label{Fig.2}
    \vspace{-0.5cm}
\end{figure*}
The asymptotic BEP for large values of $\delta$ (i.e., $\delta \to \infty$) is approximated as
\begin{align}
    &P_{\text{e}}^{\infty} = 
    \frac{1}{2} \left[ 1 - F_{{Z}} \left( \frac{2 \alpha N \ln(\alpha)}{\alpha -1} \right)  +   F_{{Z}} \left( \frac{2N \ln(\alpha)}{\alpha-1}\right)\right] \nonumber\\
    & \times
    \int_0^\infty \dfrac{r}{\sigma_\text{s}^2} I_0\left(\dfrac{r\lambda}{\sigma_\text{s}^2}\right)\exp\left(-\dfrac{r^2 + \lambda^2}{2\sigma_\text{s}^2}\right) \text{d}r \nonumber\\
    & = \frac{1}{2} \left[ 1 - F_{{Z}} \left( \frac{2 \alpha N \ln(\alpha)}{\alpha -1} \right)  +   F_{{Z}} \left( \frac{2N \ln(\alpha)}{\alpha-1}\right)\right] \nonumber\\
    & \times \left( 1 - Q_{1}(\lambda/\sigma_\text{s}, \infty) \right) \nonumber\\
    & = \frac{1}{2} \left[ 1 - F_{{Z}} \left( \frac{2 \alpha N \ln(\alpha)}{\alpha -1} \right)  +   F_{{Z}} \left( \frac{2N \ln(\alpha)}{\alpha-1}\right)\right].
\label{eq:asymptotic}
\end{align}

As \eqref{eq:asymptotic} demonstrates, the BEP forms an error floor by being independent of $\delta$ and $|h|$. This means that even with an improved useful noise variance and perfect CSI, the BEP will not decrease below the floor.

\subsection{Approximate BEP}

The approximate BEP of the TNC system for $h$ constant, based on the $Q$-function and valid only for large $N$, is given by \cite[Eq. (14)]{basar2024noise}.
Therefore, using that equation and following the same rationale used to find \eqref{eq: BEP Rice def}, then the approximate BEP for Rician fading is given by
\begin{equation}
\begin{split}
    \tilde{P}_\text{e} &= 
\int_{0}^{\infty}{Q\left( \frac{\sqrt{N} r^2 \delta (\alpha - 1)}{2 + r^2 \delta(\alpha + 1)} \right)}\frac{r I_0\left( \frac{\lambda r }{\sigma_\text{s}^2} \right)}{\sigma_\text{s}^2 \exp\left(\frac{r^2 + \lambda^2}{2\sigma_\text{s}^2} \right)}  \text{d}r.
\end{split}\label{eq:27}
\end{equation}

\section{Simulation Results and Discussion}\label{Sec.IV}

This section compares the theoretical and simulated results of the TNC system considering two-channel scenarios, namely constant and Rician fading channels, under various configurations.
We assess the BEP derived using the exact PDF of $\hat{\sigma}_{x}^2$ and the consequent optimal bit threshold, \( \gamma(|h|)\), which is henceforth defined as \( \gamma_{\text{opt}}\), against the BEP derived using the Gaussian approximation of $\hat{\sigma}_{x}^2$ and the consequent sub-optimal bit detection threshold, \(\gamma_{\text{subopt}}\), as presented in \cite{basar2023communication} and \cite{basar2024noise}. 
The simulated BEP is achieved through Monte Carlo simulations.
\begin{figure}[b!]
    \vspace{-0.7cm}
    \centering         \includegraphics[width=0.9\linewidth]{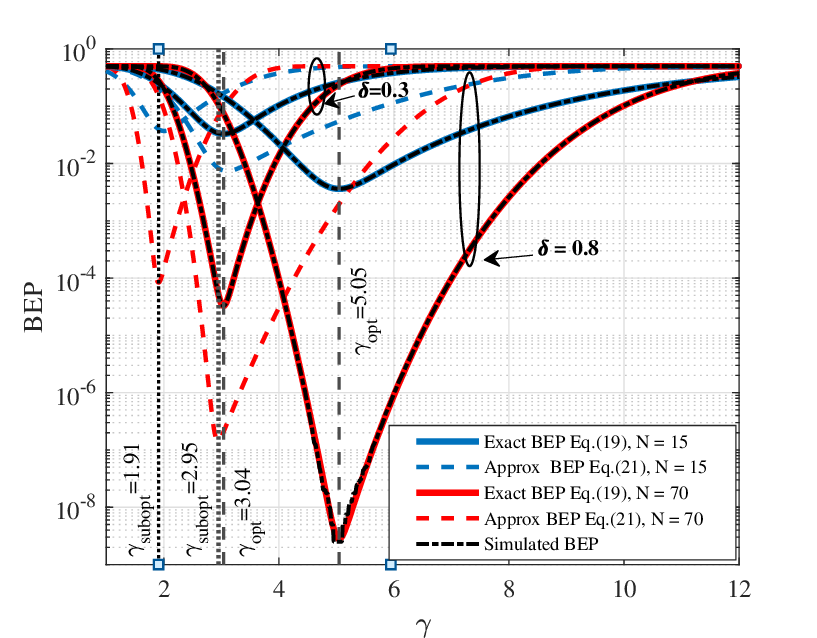}
         \vspace{-0.3cm}
         \caption{Exact, approximate, and simulated BEP curves as a function of \(\gamma\) for $\delta = \{0.3, 0.8\}$ and \(N = \{15, 70\}\), with \(\alpha = 8\) and \(K = 3\). The vertical lines present the optimal and suboptimal thresholds.}
    \label{fig: TNC-R}
\end{figure}

Fig.~\ref{Fig.2} presents BEP curves under constant channel fading (i.e., \(h = 1\)). 
The effect of \(\alpha=\{8, 25\}\) is presented in Fig.~\ref{Fig.2}(a),  with fixed \(\delta=0.3\) and \(N=30\), while Fig.~\ref{Fig.2}(b) shows the effect of \(N=\{15, 50\}\) and $\alpha=8$, with different values of \(\delta=\{0.3, 0.8\}\).
It is observed in Fig.~\ref{Fig.2}(a) that the results demonstrate a noticeable gap between the exact BEP derived in \eqref{eq15} and the approximated BEP presented in \cite[eq.~(8)]{basar2024noise}. 
This deviation becomes more pronounced as either \(\delta\) and/or \(N\) increases.
It is clear that the derived optimal bit detection threshold \(\gamma_{\text{opt}}\) provides a significant BEP improvement compared to the suboptimal detection threshold \(\gamma_{\text{subopt}}\), particularly for higher values of $\delta$, as shown in Fig.~\ref{Fig.2}(b).
For instance, the BEP decreases from approximately \(3.74\times10^{-4}\) to \(4.77\times10^{-7}\)  when employing 
\(\gamma_{\text{opt}}\). 
Moreover, increasing $N$ leads to a smaller discrepancy between the exact and approximated BEP for both noise levels (i.e., \(\delta=0.3\) and \(\delta=0.8\)) under a constant channel condition. 
The results also indicate a close match between the simulated and the exact BEP curves. 
Fig.~\ref{Fig.2}(c) presents BEP curves as a function of $\delta$, considering $\alpha=8$ and $N = \{ 15, 30, 50\}$.
Note that as \(N\) increases, a high performance in terms of BEP is achieved.
For \(\delta=30\) and $N=50$, the BEP reduces to \(5.26\times10^{-10}\), confirming the strong influence of the sample size, \(N\), on the performance.
\begin{figure*}[ht]
    \centering
    \subfloat[]{
        \includegraphics[trim=0cm 0cm 0cm 0cm, clip, width=0.325\textwidth]{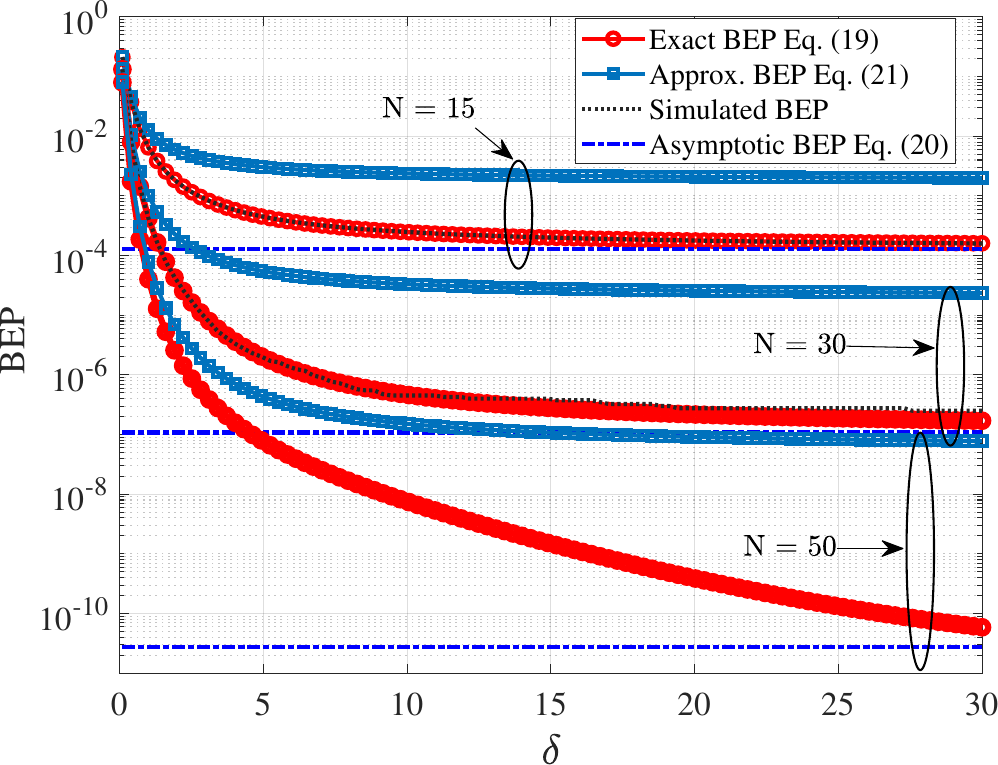}
        \label{fig:4a}
    }
    \subfloat[]{
        \includegraphics[trim=0cm 0cm 0cm 0cm, clip, width=0.325\textwidth]{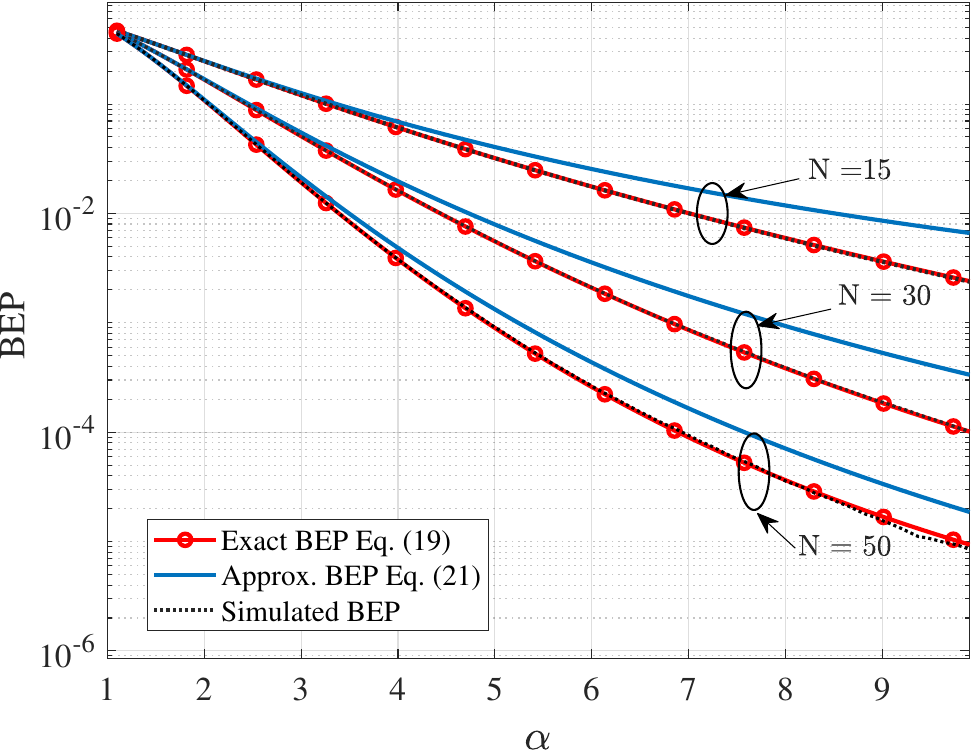}
        \label{fig:4b}
    }
    \subfloat[]{
        \includegraphics[trim=0cm 0cm 0cm 0cm, clip, width=0.325\textwidth]{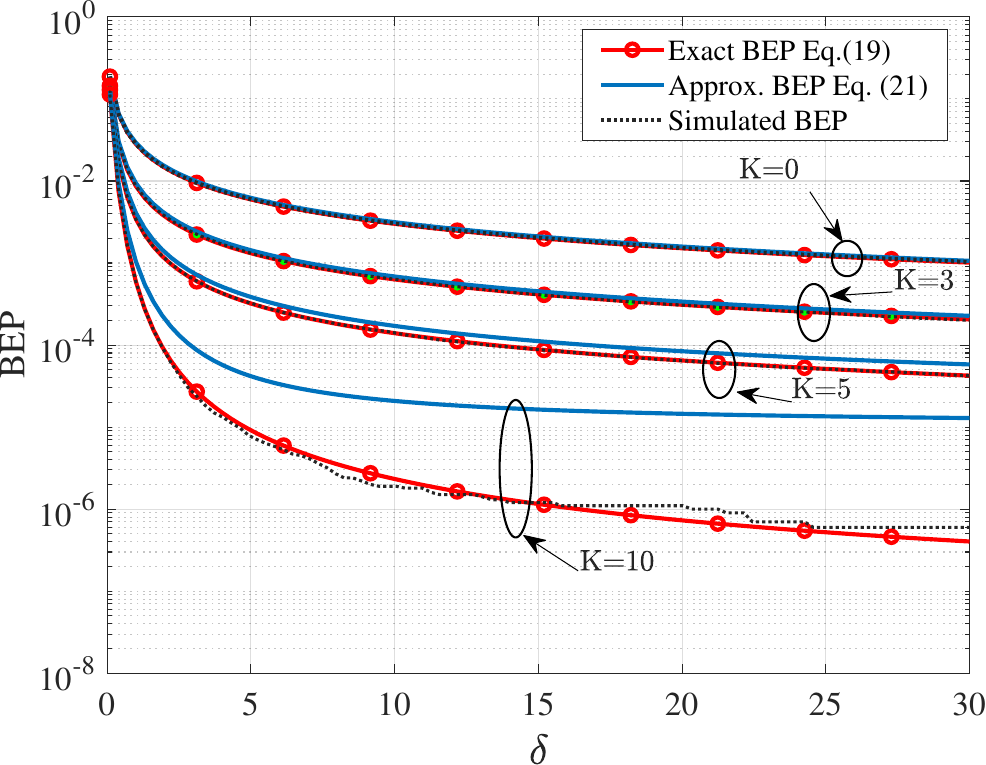}
        \label{fig:4c}
    }
    \caption{BEP under Rician fading:  
    (a) versus $\delta$ for $\alpha=8$, $K=10$, and varying $N$;  
    (b) versus $\alpha$ for $\delta=0.8$, $K=3$, and varying $N$;  
    (c) versus $\delta$ for $\alpha=8$, $N=30$, and varying $K$.}
    \label{Fig.4}
    \vspace{-0.7cm}
\end{figure*}

Fig.~\ref{fig: TNC-R} depicts the BEP as a function of the detection threshold $\gamma$ under Rician fading conditions.
It considers $N = \{15, 70\}$, $\delta = \{0.3, 0.8\}$, $\alpha = 8$, $K = 3$, and $N_a=30$.
As expected, increasing the number of samples, $N$, substantially enhances the BEP performance, with additional gains observed for higher values of $\delta$. 
As also shown in Fig.~\ref{fig: TNC-R}, the BEP decreases significantly from $3.28\times10^{-5}$ to $2.80\times10^{-9}$ when $\delta$ increases from $0.3$ to $0.8$ and $N$ increases from $15$ to $70$, respectively. 
The results also demonstrate a noticeable gap between the exact and approximated BEP curves derived in~\eqref{eq25} and \eqref{eq:27}, respectively. 
Moreover, the results show a strong agreement between the simulated and exact BEP curves. 
Additionally, it is noticeable that the difference between exact and approximated BEP curves increases with $\delta$. 
The use of the proposed optimal detection threshold, \( \gamma_{\text{opt}}\), notably enhances the achieved BEP performance compared to \( \gamma_{\text{subopt}}\), with the improvement becoming increasingly evident as $\delta$ increases. 
Furthermore, increasing the value of $N$ can compensate for higher noise levels (i.e., smaller $\delta$ values). 
This trend is demonstrated by the enhanced performance achieved in high-noise conditions (for instance, $\delta = 0.3$) with $N = 70$, relative to lower-noise cases (for instance, $\delta=0.8$) with $N = 15$. 
As depicted in Fig.~\ref{Fig.4}(a), larger values of $\delta$ and $N$ lead to a noticeable improvement in BEP performance.
Since $\delta$ can be interpreted as the SNR in conventional communication systems, it explains the observed improvement in BEP performance as it increases. 
Additionally, as $\delta$ increases, the BEP curves touch the asymptotic lines. Moreover, as expected, the BEP reduces as $N$ increases. 
This happens because the variance estimate, $\hat{\sigma}_{x}^2$, becomes more precise with $N$. 
The impact on the BER caused by varying \(\alpha\) and \(K\) is shown in Fig.~\ref{Fig.4}(b) and (c), respectively.
Remarkably, increasing either one of the parameters reduces the BEP. 
As expected and depicted in Fig.~\ref{Fig.4}(b), by increasing \(\alpha\), the distance between the variances \(\sigma_0^2\) and \(\sigma_1^2\) increases, which enhances detection accuracy, reducing the BEP. 
It is also worth noting that \( \gamma_{\text{opt}}\) attains lower BEP than \( \gamma_{\text{subopt}}\), particularly at high $\alpha$ regime. However, the gap between the exact and approximated BEP curves decreases as $N$ increases.

Fig. \ref{Fig.4}(c) illustrates how the Rician factor, $K$, influences the BEP performance. 
As foreseen, as $K$ increases, which corresponds to a stronger LoS contribution, the BEP improves. 
However, the gap between exact and approximated curves grows larger as $\delta$ and $K$ increase. 
For $K = 0$ (i.e., Rayleigh fading), the BEP is higher compared to the cases with Rician factors \(K=3, 5\) and \(10\). 
With $K = 10$, the system delivers good reliability even for low $\delta$ values, highlighting the benefits of a dominant LoS path. 
{Specifically, for $\delta = 30$ and $K = 10$, the system achieves a BEP of approximately $4 \times 10^{-7}$, outperforming the BEPs for other values of $K$ by about 10 times.}
\section{Conclusions}\label{Sec.V}

This work has presented an accurate analytical framework for TNC systems operating under Rician fading. 
By deriving the optimal MLD threshold and a BEP expression based on chi-squared statistics and Gauss–Laguerre quadrature, the study has overcome the inaccuracies of Gaussian-based approximations.
Simulation results have confirmed the validity of the proposed model and revealed that increasing the sample size, resistance ratio, and signal-to-noise variance ratio significantly enhances detection performance. 
The proposed approach establishes a rigorous foundation for optimizing TNC systems in energy-constrained IoT and future B5G/6G scenarios, paving the way for extensions to multi-user, dynamic fading, and experimental implementations.

%
\ifCLASSOPTIONcaptionsoff
  \newpage
\fi
\bibliographystyle{IEEEtran}
\bibliography{IEEEabrv,bibliography}

\newpage

\vfill

\end{document}